\documentclass{article}
\usepackage{ijcai23}

\usepackage{times}
\usepackage{soul}
\usepackage{url}
\usepackage[hidelinks]{hyperref}
\usepackage[utf8]{inputenc}
\usepackage[small]{caption}
\usepackage{graphicx}
\usepackage{amsmath}
\usepackage{amsthm}
\usepackage{booktabs}
\usepackage{algorithm}
\usepackage{algorithmic}
\usepackage[switch]{lineno}
\usepackage{amssymb}
\newtheorem{myDef}{Definition}
\usepackage[table,xcdraw]{xcolor}
\usepackage{booktabs}
\usepackage{multirow}
\usepackage{graphicx}
\usepackage{subfigure}

\title{Inducing Stackelberg Equilibrium through Spatio-Temporal Sequential Decision-Making in Multi-Agent Reinforcement Learning}

\author{
Bin Zhang$^{1,2}$
\and
Lijuan Li$^{1,2,*}$\and
Zhiwei Xu$^{1,2}$\and
Dapeng Li$^{1,2}$\and
Guoliang Fan$^{1,2,}$\thanks{corresponding author}
\affiliations
$^1$Institute of Automation, Chinese Academy of Sciences\\
$^2$School of Artificial Intelligence, University of Chinese Academy of Sciences
\emails
\{zhangbin2020, lijuan.li, xuzhiwei2019, lidapeng2020, guoliang.fan\}@ia.ac.cn
}

\begin{document}

\maketitle

\begin{abstract}

In multi-agent reinforcement learning (MARL), self-interested agents attempt to establish equilibrium and achieve coordination depending on game structure. However, existing MARL approaches are mostly bound by the simultaneous actions of all agents in the Markov game (MG) framework, and few works consider the formation of equilibrium strategies via asynchronous action coordination. In view of the advantages of Stackelberg equilibrium (SE) over Nash equilibrium,
we construct a spatio-temporal sequential decision-making structure derived from the MG and propose an N-level policy model based on a conditional hypernetwork shared by all agents.
This approach allows for asymmetric training with symmetric execution, with each agent responding optimally conditioned on the decisions made by superior agents. 
Experiments demonstrate that our method effectively converges to the SE policies in repeated matrix game scenarios,
and performs admirably in immensely complex settings including cooperative tasks and mixed tasks.
\end{abstract}

\section{Introduction}

Reinforcement learning (RL) is a popular method for solving sequential decision-making problems, but it faces several challenges when applied to a multi-agent system (MAS). 
In a constantly changing environment, agents' rewards are often closely tied to the actions of others. 
No individual exists in isolation and their actions are often interdependent.
To achieve optimal outcomes, they must learn to find a stable equilibrium rather than simply maximizing their own returns.
It is,
therefore,
an eternal topic in multi-agent reinforcement learning (MARL) to learn how to coordinate among agents to achieve the optimal joint policy.

Currently, the majority of prevalent MARL approaches adhere to the centralized training with decentralized execution (CTDE) architecture~\cite{CTDE}.
In this paradigm, agent's policy model is trained centrally without communication constraints but executed in a distributed manner.
Nevertheless, current CTDE approaches prioritize the acquisition of comprehensive cognition about the environment and the formulation of cooperative policies by directly getting the global state~\cite{QMIX} or the local observations and actions of all agents~\cite{MADDPG}.
Some works based on game abstraction~\cite{GA2} \nocite{GA1,GA3,GA4} model the relationship of agents based on their observational data, which are state representation learning essentially. 
However, centralized training is not leveraged to its full potential.
Actually, it is feasible to construct an interaction mechanism that effectively guides agents towards mandatory coordination.

Moreover, game theory offers the fundamental framework for the interaction of multiple intelligent agents, 
and several MARL algorithms, such as Nash Q-Learning \cite{NASH-Q}, Mean Field Q-learning \cite{MFQ} and HATRPO \cite{HATRPO}, seek for convergence to Nash Equilibrium (NE).
However, it is worth noting that many game problems feature more than one NE, and the diverse NE policies chosen by different individuals frequently lead to undesirable outcomes.
In addition, in a two-player zero-sum game, the MaxMin operator~\cite{minimax_Q} can be used to calculate NE strategies for each agent, such that neither player can gain an advantage by deviating from their chosen strategy.
But in more complex general-sum game situations, finding NE strategies is typically rather difficult.
Consequently, we intend to concentrate on the Stackelberg equilibrium (SE) \cite{Stackelberg}, in which agents make decisions in a leader-follower framework, with leaders prioritizing decision-making and enforcing their policies on followers who respond rationally to this enforcement.
It has been shown that SE is a superior convergence objective for MARL compared to NE, whether from the certainty of equilibrium solution or Pareto superiority~\cite{BiRL}.
Actually, when SE encounters MARL, we aim to address the following challenges: (a) How to converge to SE policies that rely on asynchronous action coordination under the Markov game framework where agents act simultaneously? (b) How to extend the method to scenarios with more than two agents?

To address the aforementioned issues, we draw inspiration from techniques used in single-agent RL for high-dimensional continuous control tasks~\cite{SQL}. Utilizing significant benefits of centralized training, we model MARL as a sequential decision-making problem in both temporal and spatial dimensions, and then train the heterogeneous SE policies asymmetrically.
Furthermore, to facilitate the execution of SE policies in a communication-free and symmetrical environment, we introduce an N-level policy model based on a conditional hypernetwork shared by all agents, with synthetic targets of generating the weights of target agents' policy networks according to their priority attributes. 
This approach allows us to bridge the gap between environmental communication restrictions and the requirements that SE policy needs to access superior agents' policies.
It mitigates the problem of suboptimal solutions induced by parameter sharing, and overcomes the issue of learning and storage costs caused by heterogeneous policy learning.
Our primary contributions are summarized as follows:
\begin{itemize}
    \item We develop a spatio-temporal sequential Markov game framework based on agent priority that enables agents to establish an efficient interaction mechanism during centralized training.
    \item We construct an N-level policy model to assist agents in executing SE policies in a fully decentralized setting without imposing a limit on the number of agents.
    \item We establish the asymmetric training with symmetric execution  paradigm, a significant augmentation of CTDE.
    \item Our method is demonstrated to converge to the SE through repeated matrix game experiments, and the results in more complicated scenarios also illustrate its superiority over powerful benchmarks in terms of sample efficiency and overall performance.
\end{itemize}

\section{Preliminaries}

\subsection{Markov Game}
\label{Markov Game}

The multi-agent decision-making problem is typically described as a Markov game (MG), which can be defined by a tuple $\Gamma \triangleq \langle \mathcal{I}, \mathcal{S}, \{\mathcal{A}^i\}_{i\in\mathcal{I}}, P, \{r^i\}_{i\in\mathcal{I}}, \gamma\rangle$.
$\mathcal{I}=\{1,2,...,n\}$ denotes the set of agents and $s\in S$ is the global state set of the environment.
$\mathcal{A}^i$ denotes the action space of agent $i$ and the joint action space $\mathcal{A}=\prod_{i=1}^{n}\mathcal{A}^i$ is the product of action spaces of all agents.
$P:\mathcal{S\times \mathcal{A}}\to \Omega(\mathcal{S})$ denotes state transition function, where $\Omega(X)$ is the set of probability distributions in $X$ space.
$r^i:\mathcal{S\times \mathcal{A}\to \mathbb{R}}$ is the reward function of agent $i$
and $\gamma$ is the discount factor. 
At time step $t$,  each agent chooses an action $a^i_t\in\mathcal{A}^i$ at state $s_t\in\mathcal{S}$ based on its own policy $\pi^i:\mathcal{S}\to\Omega(\mathcal{A}^i)$ and receives feedback in the form of $r^i(s_t, \boldsymbol{a_t})$, 
where $\boldsymbol{a_t}=(a_t^1,...,a_t^n)\in \mathcal{A}$.
The environment moves to a new state $s_{t+1}\sim P(s_{t+1}\mid s_t,a_t)$ as a result of joint action $\boldsymbol{a_t}$.
The joint policy of all agents is expressed as $\boldsymbol{\pi}\left(s_{t}\right)=\prod_{i=1}^{n} \pi^{i}\left(s_{t}\right)$.
Under the framework of MG, the state value function of agent $i$ is defined as :
\begin{equation}
\resizebox{.91\linewidth}{!}{$
\displaystyle
V^i_{\boldsymbol{\pi}}(s) = \mathbb{E}_{s\sim P,a^{-i} \sim \boldsymbol{\pi}^{-i}}\left[\sum_{t=0}^{\infty} \gamma^{t} r^i_{t}(s_t,\boldsymbol{a_t}) \mid \mathrm{s}_{0}=s,a_t^i\sim\pi^i(s_t)\right],
$}
\end{equation}
where $-i$ represents all agents except $i$. According to Bellman equation, the action-state value function is denoted as:
\begin{equation}
\begin{aligned}
Q^i_{\boldsymbol{\pi}}(s, \mathbf{a})=r^i(s,\mathbf{a})+\gamma\sum_{s'\in \mathcal{S}}P(s'|s,\mathbf{a})\cdot V^i_{\boldsymbol{\pi}}(s').
\end{aligned}
\label{MG-Q}
\end{equation}

\subsection{Multi-Agent Reinforcement Learning}
\begin{figure}[t]
    \centering
    \includegraphics[width=3.1 in]{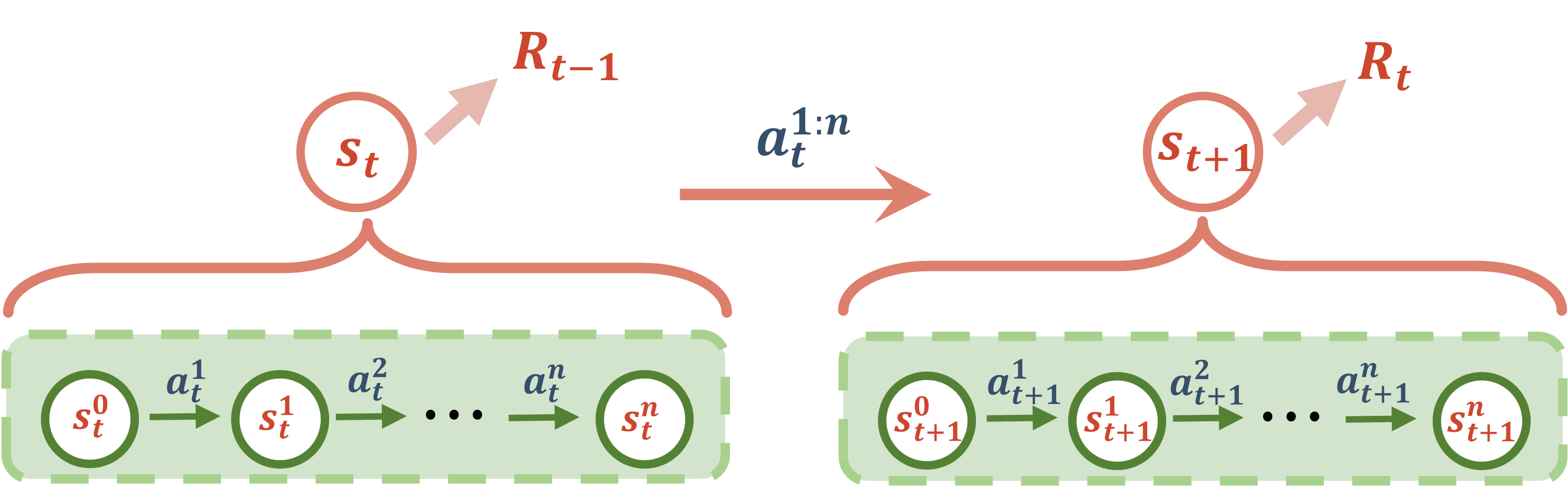}
    \caption{The STMG state transition procedure. It is an extensive game version of MG, which specifies the decision-making sequence of agents simultaneously.}
    \label{fig:EMG}
\end{figure}
In a multi-agent system, each decision-maker strives to maximize their own expected utility, denoted as $\mathcal{J}^i(\theta)=\mathbb{E}_{s\sim P,a \sim \pi_\theta}\left[\sum_{t=0}^{\infty} \gamma^{t} r^i_{t} \right]$, where $\theta=\{\theta^i\}_{i\in{\mathcal{I}}}$ represents the policy parameters of agents.
However, using gradient ascent to directly update the strategy, i.e., $\theta^i\leftarrow \theta^i+\alpha\nabla_{\theta^i}\mathcal{J}^i(\theta)$, usually fails because $\mathcal{J}^i(\theta)$ is affected by all agents and the gradient update directions among agents may conflict.
Consequently, it is vital to employ the concept of game equilibrium solutions in order to develop effective coordination strategies. 
One common solution objective is NE. 
Nash-Q learning, for example, computes agent $i$'s value function $V_{\mathcal{N} a s h}^i(s)$ when all agents follow the NE strategy in each stage of the game, 
and updates the action-state value function $Q_{\mathcal{N} a s h}^i(s, \mathbf{a})$ through Eq.~\eqref{MG-Q}. 
The algorithm can converge to NE strategies under the strong assumption that equilibrium exists at every game stage.
All agents adopt equilibrium policies as their own convergence objective rather than selfishly maximizing their own individual utilities.
With the emergence of deep learning, current popular MARL approaches frequently employ the CTDE paradigm to enable coordination via credit assignment~\cite{QMIX} or centralized critics~\cite{MADDPG}.

\begin{figure*}[htbp]
    \centering
    \includegraphics[width=5.4 in]{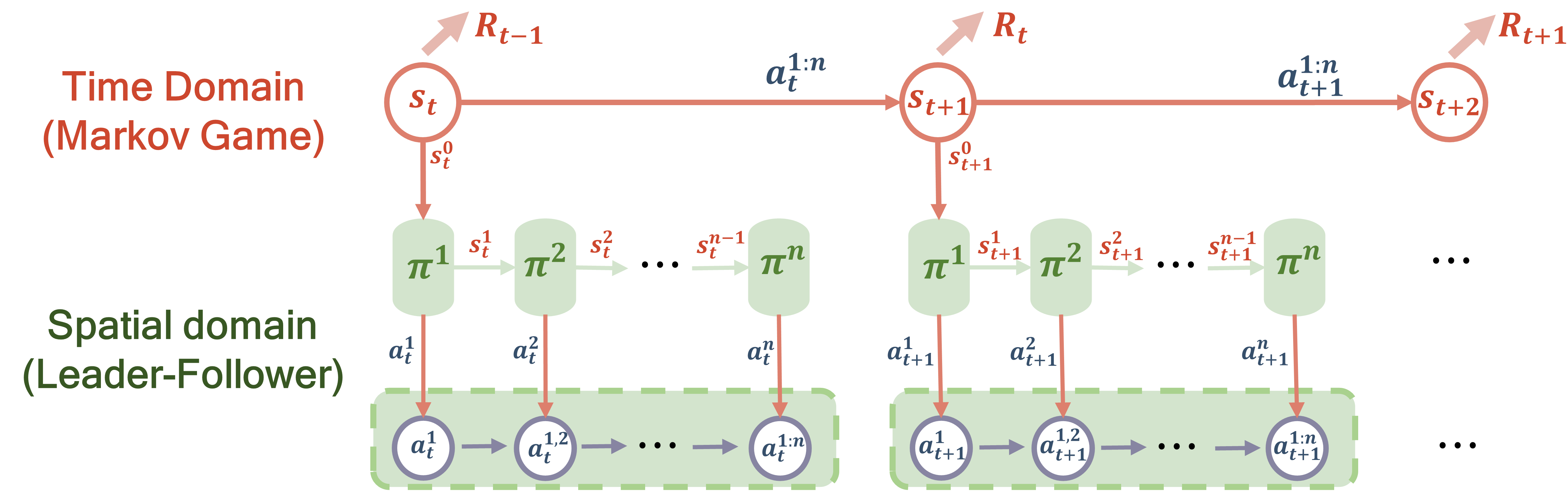}
    \caption{Exemplification of the execution mechanism under the STMG framework. In the temporal domain, all agents continue to adhere to the MG settings. Simultaneously, agents make decisions according to the Stackelberg leadership model, and subgame states $s^{i-1}_t$ are introduced to maintain each agent's policy inputs.}
    \label{fig:framework}
\end{figure*}
\subsection{Stackelberg Leadership Model}

The Stackelberg leadership model dictates the order of actions among agents. Consider a two-player game, the leader can enforce his own strategy $\pi^1$ to the follower, who will react after observing the leader's behavior $\pi^2:\mathcal{S\times A}^1\to\Omega(\mathcal{A}^2)$, where $\pi^i \in \Pi^i$ and $\Pi^i$ represents the policy space.
In this model,
leader optimizes its objective on the premise that follower will provide the best response $BR(\pi^1)$, while the follower tends to maximize its expected utility based on leader's preconditions.
It can be formalized as:
\begin{equation}
\begin{aligned}
\max_{\pi^1 \in \Pi^1} \{\mathcal{J}^1(\pi^1, \pi^2)|\pi^2&=BR(\pi^1)\},\\
\max_{\pi^2 \in \Pi^2}\{\mathcal{J}^2(a^1, \pi^2)|&a^1 \sim \pi^1\}.
\end{aligned}
\label{SLM}
\end{equation}
Equilibrium signifies that in a multi-player game, all players have adopted the optimal strategy and none can improve their performance by altering their own strategy.
Correspondingly, Stackelberg equilibrium signifies that both agents adhere to the optimal solution strategies $({\pi^1_{SE}}, \pi^2_{SE})$ of the the above optimization issue.
 It satisfies:
\begin{equation}
\begin{aligned}
V^1_{\pi^1_{SE},BR(\pi^1_{SE})}(s) &\geq V^1_{\pi^1,BR(\pi^1)}(s),\\
\pi^2_{SE}&=BR(\pi^1_{SE}).
\end{aligned}
\end{equation}
As a mandatory equilibrium, SE offers more benefits than NE in terms of the stability, certainty of equilibrium points and Pareto superiority \cite{BiRL}.
Consequently, it is a more suitable learning objective.

\begin{figure*}[htbp]
    \centering
    \includegraphics[width=6.5 in]{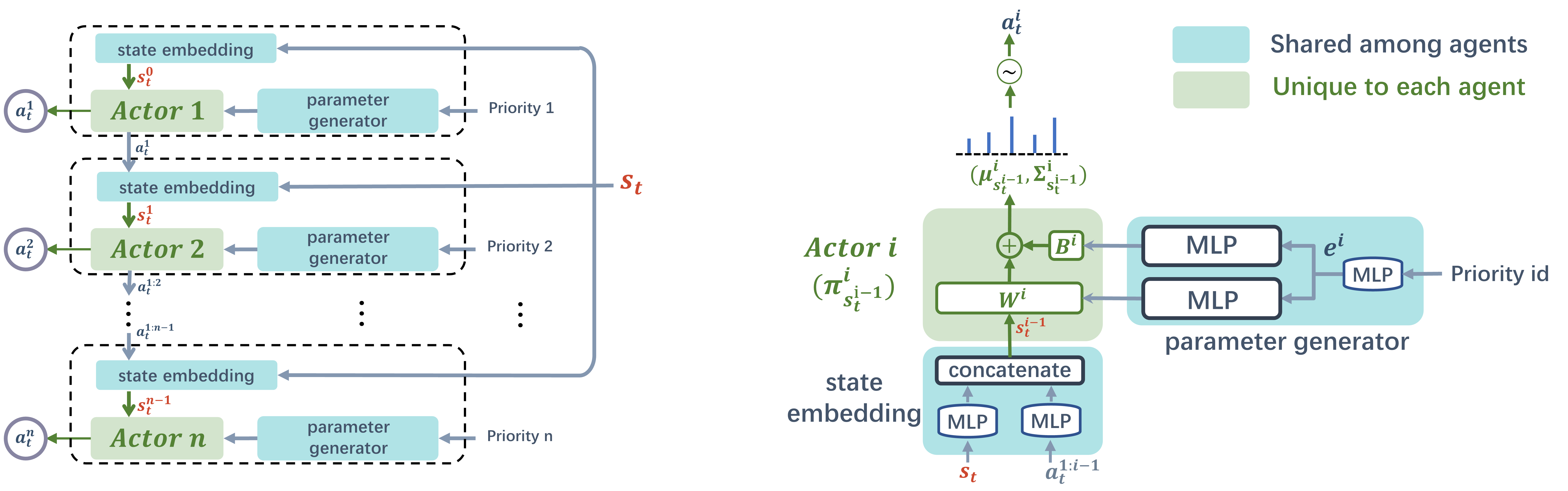}
    \caption{The overall architecture of STEP. \emph{Left}: The workflow of STEP for a comprehensive decision in a time step. Agents make their decisions based on the current situation $s_t$, their self-positioning $Priority \text{ ID}$, and the prerequisite actions $a^{1:i-1}_t$ of superior agents. \emph{Right}: The structure of N-level policy model. It allows for the implementation of heterogeneous policies under parameter sharing and the Stackelberg equilibrium policies under symmetric conditions.}
    \label{fig:stru}
\end{figure*}

\section{Methodology}

In this section, we propose a new method called \textbf{S}patio-\textbf{T}emporal sequence \textbf{E}quilibrium \textbf{P}olicy optimization (STEP). 
This method leverages the advantages of centralized training to establish an interaction mechanism and employs a more efficient policy network to facilitate the achievement of coordinate policies during decentralized execution. The following parts provide a detailed description of the procedure for implementing STEP.

\subsection{Spatio-Temporal Sequential Markov Game}


Current CTDE approaches, especially policy-based methods, tend to center around enhancing the richness of input information rather than boosting the effectiveness of collaboration.
Actually, centralized training liberates us from restrictive circumstances.
With centralized training, we can definitely create an interaction mechanism among agents for completing their tasks. 
Furthermore, the concept of the SE is more in line with the goals of RL as it requires merely additional constraints, which can be easily achieved by setting a specific interaction mechanism.
So naturally, we utilize the form of multi-player Stackelberg leadership model and construct the spatio-temporal sequential Markov game (STMG) in the training phase to direct agents coordination and  facilitate the implementation of SE policies.

\begin{myDef}
    STMG can be formalized as a tuple $\langle\mathcal{I}, \mathcal{S}, \{\mathcal{A}^i\}_{i\in\mathcal{I}}, P, \{\mathcal{r}^i\}_{i\in\mathcal{I}}, \gamma, \{{o}^i\}_{i\in \mathcal{I}}\rangle $.
In addition to the MG defined in Section \ref{Markov Game}, STMG add the term $o^i$, which denotes the action order of agent $i$. $\mathcal{O}=(o^1,...,o^n)$ represents the action order of all agents, indicating the priority/importance of agents at the decision-making stage. 
\end{myDef}
Figure~\ref{fig:EMG} shows the state transition process in STMG. For the sake of simplicity, we assume that the agent ID $i$ is assigned by priority $o^i$.
Compared with MG, STMG assumes the form of a sequence decision in both temporal and spatial domains.
Agents with higher priorities have greater initiative, whereas agents with lower priorities must respond to the actions of those with higher priorities.
Correspondingly, the policy of each agent changes to $\pi^i:\mathcal{S}\times \mathcal{A}^1\cdot\cdot\cdot \mathcal{A}^{i-1}\to\Omega(\mathcal{A}^i)$.
The action-state value function of agent $i$ can be written as:
\begin{equation}
\resizebox{.91\linewidth}{!}{$
\displaystyle
\begin{aligned}
Q^i_{\boldsymbol{\pi}}&(s,  a^{1:i-1},a^i)=\mathbb{E}_{s \sim P, a^{i+1:n} \sim \boldsymbol{\pi}^{i+1:n}}\bigg[\sum_{t=0}^{\infty} \gamma^{t} \cdot \\
&r^i_{t}(s_t,\boldsymbol{a_t})\mid \mathrm{s}_{0}=s,\mathbf{a}_{0}=\boldsymbol{a},a_t^i\sim\pi^i(s_t,a_t^1,...,a_t^{i-1}) \bigg]. 
\end{aligned}
$}
\label{STMG-Q}
\end{equation}
The state value function is denoted as:
\begin{equation}
V^i_{\boldsymbol{\pi}}(s,a^{1:i-1})=\sum_{a^i\in\mathcal{A}^i}\pi^i(a^i|s,a^{1:i-1})Q^i_{\boldsymbol{\pi}}(s,a^{1:i-1},a^i).
\label{STMG-V}
\end{equation}
And then we have the advantage function:
\begin{equation}
\resizebox{.91\linewidth}{!}{$
\displaystyle
\begin{aligned}
A^i_\pi(s,  a^{1:i-1},a^i)=Q^i_{\boldsymbol{\pi}}(s,  a^{1:i-1},a^i)-V^i_{\boldsymbol{\pi}}(s,a^{1:i-1}).
\end{aligned}
$}
\label{STMG-A}
\end{equation}

Within the STMG framework, agent with priority $i$ can consider the actions of the preceding $i-1$ superior agents as prerequisites,
maximize its own expected return $\mathcal{J}^i(\theta)=\mathbb{E}_{s\sim P,a^{i+1:n} \sim \boldsymbol{\pi}^{i+1:n}_\theta}\left[\sum_{t=0}^{\infty} \gamma^{t} r^i_{t} \right]$ according to the current subgame state $s^{i-1}=(s,a^1,...,a^{i-1})$, which is equivalent to learning the best response to the strategies of superior agents. In summary, during the training phase, agents' execution mechanism can be divided into the Markov game process in the time domain and the Leader-Follower model in the space domain. The specific execution process is depicted in Figure~\ref{fig:framework}.

\subsection{Parameterized N-level Policy Model Using A Conditional Hypernetwork}

Within the STMG framework, agents are able to acquire the actions of superior agents directly. This asymmetric approach  arises naturally in the centralized training phase without communication restrictions, but it is obviously illegal in the decentralized execution phase.
To this end, several approaches can be taken.  We can train a policy network that is shared by all agents, allowing them to calculate the actions of other agents using the shared parameters. 
Designing a communication module that enables superior agents to broadcast their decisions to inferior agents is also an alternative.
Moreover, agents can keep copies of other players' policy or value networks to independently calculate their own policies.
Nonetheless, parameter sharing can result in suboptimal solutions, and communication may not be feasible in fully decentralized execution settings or with communication bandwidth constraints. 
Additionally, the scalability of the algorithm may be limited as the number of agents increases and the volume of saved model copies grows.

In contrast to the aforementioned methods, we develop an N-level policy model using a conditional hypernetwork. 
It consists of a parameter generator, a state embedding, and a target policy module. 
Figure~\ref{fig:stru} illustrates its structure in detail. 
MARL is essentially a multi-task learning (multi-objective regression) problem~\cite{MAAC}.
Instead of explicitly training and storing individual policy modules for each agent, 
we maintain a set of policy parameters $\{\theta_{tar}^i\}_{i\in \mathcal{I}}$ by training a meta-model (conditional hypernetwork) $\mathcal{H}(e^i, \theta_h)=\theta^i_{tar}$ with the weights $\theta_h$ shared by all agents.
It maps the embedding of each agent's priority ID $e^i$ to the parameter configuration of its policy network.
It has been demonstrated that hypernetworks can handle continuous learning tasks~\cite{hyperCL}, where a model is trained on a series of tasks in sequence.
This is highly compatible with our expectations.
In addition, the input $s^{i-1}$ to the target policy module is encoded by the state embedding network $\mathcal{E}(s,a^1,...,a^{i-1};\mathcal{\theta}_s)$,  which receives as input the current state of the environment and the actions of superior agents.
Specifically, each component of the model is described as follows:

\vspace{-0.5cm}
\begin{align}
&\text{State Embedding:} &s^{i-1}&=\mathcal{E}(s,a^1,...,a^{i-1};\mathcal{\theta}_s),\\
&\text{Parameter Generator:} &\theta^i_{tar}&= \mathcal{H}(e^i;\theta_h),\\
&\text{Target Policy:}  &a^i&\sim\pi^i(s^{i-1};\theta^i_{tar}).
\label{STEP}
\end{align}

It is important to note that the parameters of the policy model consist of $\theta^i=(\theta_s,\theta_h,\theta^i_{tar})$, where $\theta_s,\theta_h$ are learnable parameters shared by all agents and $\theta^i_{tar}$ can be accessed by all agents through the parameter generator. By using the N-level policy model, we are able to share policy parameters while avoiding the suboptimal solutions and storage issues previously mentioned, resulting in efficient decentralized coordination.

\subsection{Implementation}
The N-level policy model allows for the implementation of the asymmetric training with symmetric/decentralized execution (ATSE) paradigm in STEP. 
Meanwhile, we select Proximal Policy Optimization (PPO)~\cite{PPO} as the underlying algorithm due to its superior performance. 
To ensure the balance between exploration and exploitation during the training phase, 
agents sample actions from the categorized distribution or multivariate Gaussian distribution generated by the policy model.
However, due to the unpredictability of sampling, agents are unable to calculate the actions of superior agents.
As a result, we directly transmit the behaviors of superior agents to inferior agents during the training process.
During the execution process, each agent selects the action with the highest probability, and inferior agents can easily compute actions of superior agents through the shared  policy model  to establish corresponding coordination.
Additionally,
throughout each epoch of the training process, 
each agent trains the identical policy module using its own data, which can result in catastrophic forgetting.
To address this issue, we incorporate a regularization item to ensure that the policy network can train the current agent's policy parameters while maintaining the capacity to fit the previously updated policies of other agents. The objective function of actor $i$ is expressed as:
\begin{equation}
\begin{aligned}
    &\mathcal{L}^i(\theta) =\mathbb{E}_{s\sim P, \mathbf{a}\sim\boldsymbol{\pi}}
    [\mathcal{L}_{clip}
    -
    \mathcal{L}_h(\theta_h,\theta_{h_{old}},e^{1:i-1})\\ 
    &\quad\quad\quad\quad\quad\quad\quad\quad+
    \eta S(\pi_\theta^i(s,a^{1:i-1}))
    ],\\
    &\mathcal{L}_{clip}=\min(r^i_\theta A^i_\pi,
    clip(r^i_\theta,1\pm \epsilon)A^i_\pi),\\
    &\mathcal{L}_h(\cdot)=\frac{\beta}{i-1} \sum_{j=1}^{i-1}||\mathcal{H}(e^j;\theta_h)-\mathcal{H}(e^j;\theta_{h_{old}})||^2_2,
    \label{L_Actor}
\end{aligned}
\end{equation}
where $S(\cdot)$ is Shannon entropy used for strengthening exploration,$\quad r^i_\theta=\frac{\pi^i_\theta(a^i|s,a^{1:i-1})}{\pi^i_{\theta_{old}}(a^i|s,a^{1:i-1})}$ is the likelihood ratio between the current and previous policies, $\epsilon$ is the clipping ratio, $\eta$ and $\beta$ are coefficients of entropy and regularization term respectively.
Critic network is used to fit the value function, and its loss function is expressed as:
\begin{equation}
\resizebox{.91\linewidth}{!}{$
\displaystyle
\begin{aligned}
    \mathcal{L}&(\phi^i)= \max \big[ \left(V_\phi\left(s, a^{1:i-1}\right)-R^i\right)^2,\\
    &\left(\operatorname{clip}\left(V_\phi\left(s, a^{1:i-1}\right),
    V_{\phi_{o l d}}\left(s, a^{1:i-1}\right)\pm\varepsilon\right)-R^i\right)^2\big] ,
    \label{L_critic}
\end{aligned}
$}
\end{equation}
where $R^i$ is the cumulative return and $\varepsilon$ is the clipping ratio.

The pseudo-code of STEP can be found in Appendix.
Notably, our method provides comprehensive advantages over currently popular MARL techniques.
Unlike prior work based on SE~\cite{AQL,BiRL}, STEP is easily extensible to include additional players; Unlike prior work based on parameter sharing~\cite{QMIX}, STEP is able to learn heterogeneous policies; In contrast to previous work on learning heterogeneous policies~\cite{HATRPO}, STEP does not increase the learning cost as the number of agents increases; In contrast to previous CTDE work~\cite{MADDPG,MAPPO}, we no longer focus on merely expanding the information space, but on specific forms of coordination.

\section{Related Work}

Our study, inspired by the Stackelberg leadership model, 
builds an asymmetric cooperative connection within the MG framework and applies RL to discover the Stackelberg equilibrium solution.
In certain publications, SE strategies also serve as the convergence objective.
Similar to Nash-Q learning, Asymmetric Q-learning (AQL)~\cite{AQL} updates the action-state value function in an asymmetric setting by calculating the SE of the stage game at each iteration.
Leaders are able to access followers' reward information and save copies of each follower's value function.
Bi-AC~\cite{BiRL} presents a bi-level actor-critic method based on CTDE paradigm that employs a Q-learning-based leader~\cite{DQN} and DDPG-based follower~\cite{DDPG}.
Both the leader and follower must save the leader's critic network and the follower's actor network during the execution process in order to compute and execute their policies.
Additionally, Bully~\cite{Bully} and DeDOL~\cite{DeDOL} are also dedicated to the solution of SE.
However, all of the previous works only divide the agents into two levels, and extending them to more agents is not as straightforward as they anticipated.
Furthermore, HATRPO~\cite{HATRPO} uses random sequence updating to ensure monotonic policy improvement in multi-agent settings, 
but it is only suitable for cooperative environments. 
The algorithm requires to define a joint advantage function for all agents, limiting its application to mixed tasks.

When dealing with structured combination action spaces, single-agent RL also uses similar spatial-temporal sequence learning approaches.
SQL~\cite{SQL} imitates the sequence-to-sequence model of the structured prediction problem, 
predicts the policy of each dimension simultaneously, and combines them into a comprehensive high-dimensional policy. 
Our technique can be considered as an extension of SQL for multi-agent systems.
The actions of each dimension in SQL corresponds to the actions of  each agent in STEP, 
and the complete high-dimensional policy relates to the joint policy of all agents.
In addition, BQD's~\cite{BQD} use of the action branch structure is also an effective way to decompose and separately control high-dimensional actions.
This form of single-agent algorithm lacks communication restrictions and can freely access each dimension's policy, hence reducing the complexity of problem-solving.
In MARL, however, it is required to build extra procedures to address communication restriction-related issues.

\begin{figure}[t]
    \centering
    \includegraphics[width=3.0 in]{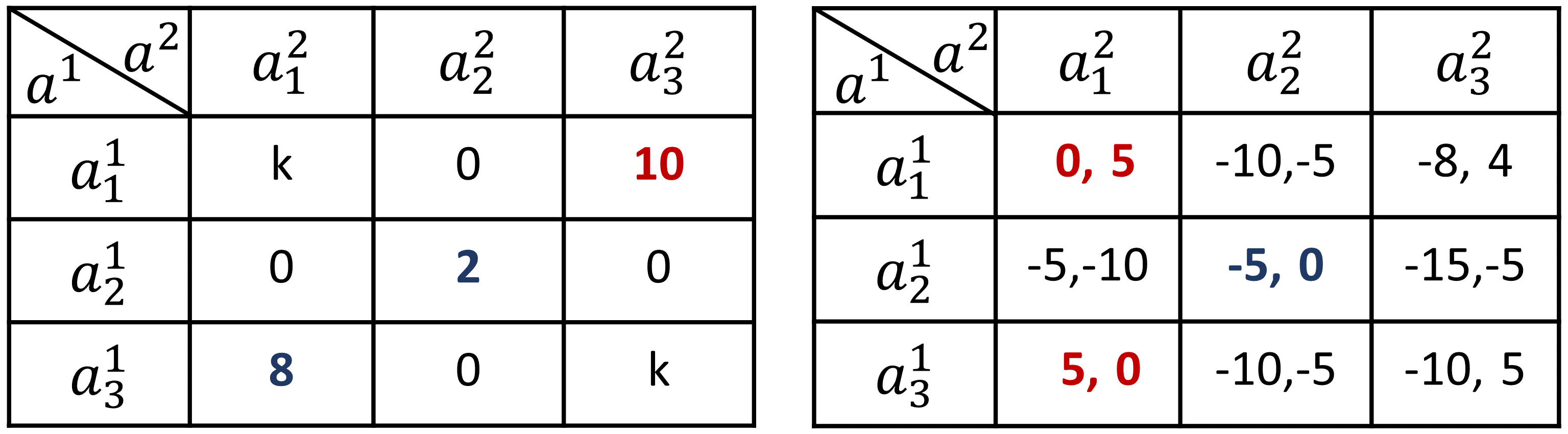}
    \caption{Matrix games. \emph{Left}: The common-payoff matrix of the Penalty game, where $k$ = $[0,-25,-50,-75,-100]$. $(a^1_1,a^2_3),(a^1_2,a^2_2)$ and $(a^1_3,a^2_1)$ are NE points, and $(a^1_1,a^2_3)$ is the only SE point. \emph{Right}: The payoff matrix of the Mixing game. It has only one NE point $(a^1_2,a^2_2)$ and only one SE point $(a^1_1, a^2_1)$. The SE is Pareto superior to the NE.}
    \label{fig:matrix}
\end{figure}

\begin{figure}[t]
    \centering
    \includegraphics[width=3.3 in]{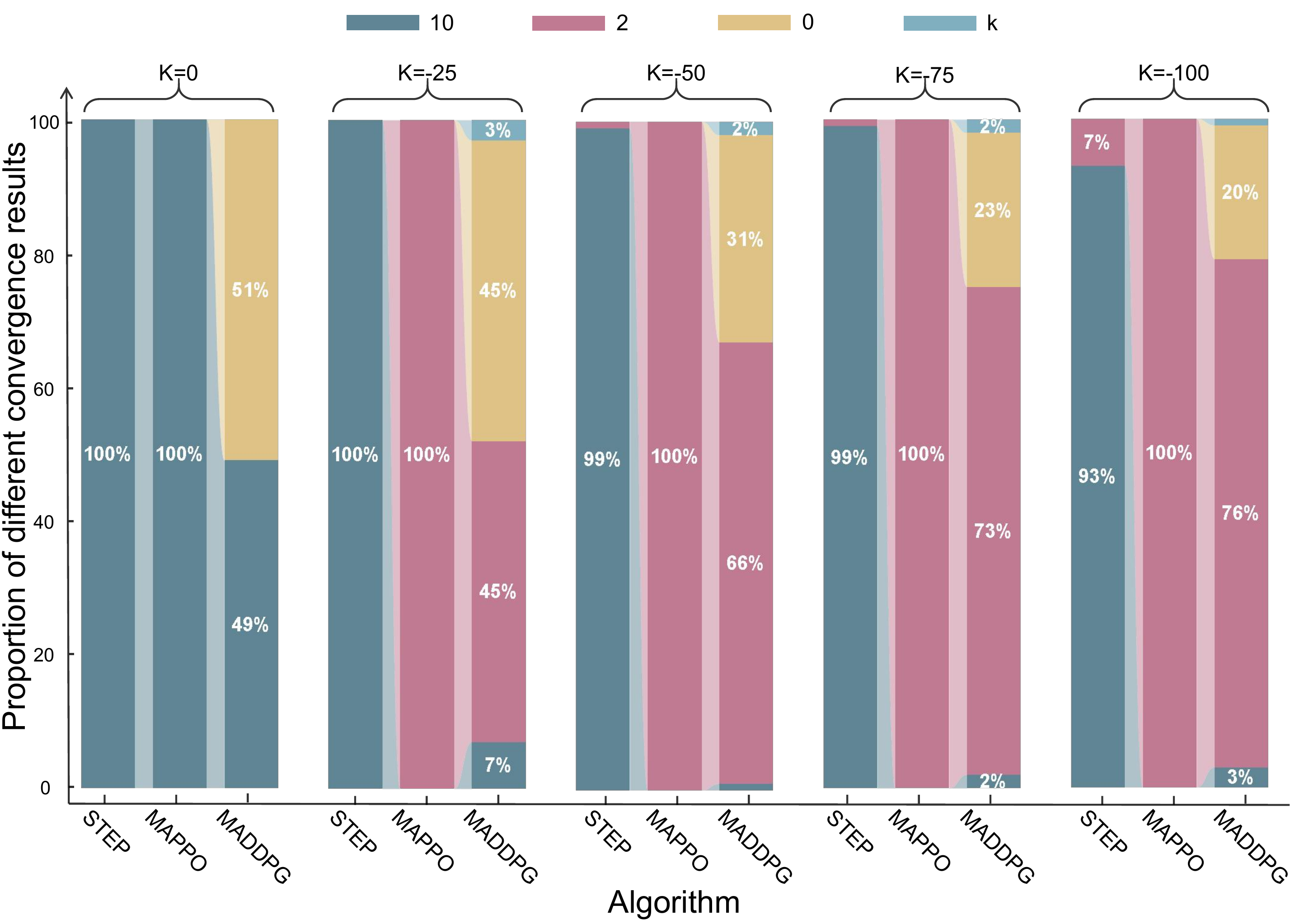}
    \caption{Proportion of different convergence results in Penalty.}
    \label{fig:penalty}
\end{figure}

\section{Experiments}

\begin{figure*}[ht]
    \centering
    \includegraphics[width=6.5 in]{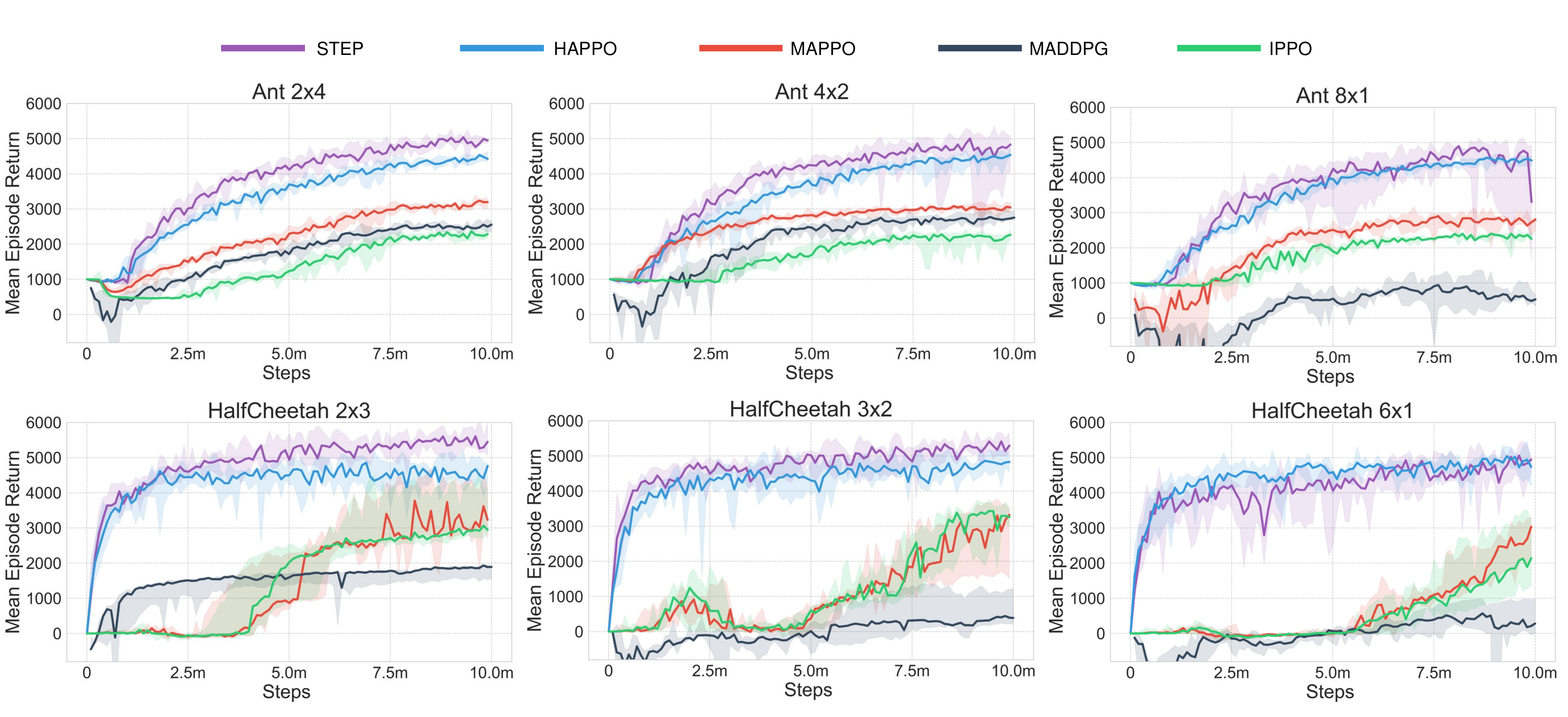}
    \vskip -0.15 in
    \caption{Performance comparison with baselines on Multi-Agent MuJoCo tasks. Error bars are a 95\% confidence interval across 5 runs.}
    \label{fig:mujoco_results}
\end{figure*}

\begin{table*}[htbp]
\centering
\resizebox{\textwidth}{!}{%
\begin{tabular}{@{}cccc|ccc|ccc@{}}
\toprule
\rowcolor[HTML]{FFFFFF} 
\cellcolor[HTML]{FFFFFF}{\color[HTML]{333333} } &
  \multicolumn{3}{c|}{\cellcolor[HTML]{FFFFFF}{\color[HTML]{333333} Agent 1}} &
  \multicolumn{3}{c|}{\cellcolor[HTML]{FFFFFF}{\color[HTML]{333333} Agent 2}} &
  \multicolumn{3}{c}{\cellcolor[HTML]{FFFFFF}{\color[HTML]{333333} Reward}} \\ \cmidrule(l){2-10} 
\rowcolor[HTML]{FFFFFF} 
\multirow{-2}{*}{\cellcolor[HTML]{FFFFFF}{\color[HTML]{333333} \textbf{}}} &
  {\color[HTML]{333333} $a^1_1$} &
  {\color[HTML]{333333} $a^1_2$} &
  {\color[HTML]{333333} $a^1_3$} &
  {\color[HTML]{333333} $a^2_1$} &
  {\color[HTML]{333333} $a^2_2$} &
  {\color[HTML]{333333} $a^2_3$} &
  {\color[HTML]{333333} agent 1} &
  {\color[HTML]{333333} agent 2} &
  {\color[HTML]{333333} Avg.} \\ \midrule
\rowcolor[HTML]{FFFFFF}
\rule{0pt}{10.6pt}
{\color[HTML]{333333} STEP} &
  {\color[HTML]{333333} $\boldsymbol{99.99_{(0.01)}}$} &
  {\color[HTML]{333333} $0.00_{(0.00)}$} &
  {\color[HTML]{333333} $0.01_{(0.01)}$} &
  {\color[HTML]{333333} $\boldsymbol{99.99_{(0.01)}}$} &
  {\color[HTML]{333333} $0.00_{(0.00)}$} &
  {\color[HTML]{333333} $0.01_{(0.01)}$} &
  {\color[HTML]{333333} $0.00_{(0.00)}$} &
  {\color[HTML]{333333} $5.00_{(0.00)}$} &
  {\color[HTML]{333333} $\boldsymbol{2.50_{(0.00)}}$} \\
\rowcolor[HTML]{FFFFFF} 
\rule{0pt}{10.6pt}
\cellcolor[HTML]{FFFFFF}{\color[HTML]{333333} MAPPO} &
  {\color[HTML]{333333} $32.93_{(46.20)}$} &
  {\color[HTML]{333333} $0.03_{(0.20)}$} &
  {\color[HTML]{333333} $67.04_{(46.21)}$} &
  {\color[HTML]{333333} $61.98_{(47.31)}$} &
  {\color[HTML]{333333} $0.04_{(0.22)}$} &
  {\color[HTML]{333333} $37.97_{(47.31)}$} &
  {\color[HTML]{333333} $2.72_{(2.38)}$} &
  {\color[HTML]{333333} $-1.29_{(6.43)}$} &
  {\color[HTML]{333333} $0.72_{(2.33)}$} \\
\rowcolor[HTML]{FFFFFF}
\rule{0pt}{10.6pt}
{\color[HTML]{333333} MADDPG} &
  {\color[HTML]{333333} $21.39_{(37.14)}$} &
  {\color[HTML]{333333} $5.62_{(20.52)}$} &
  {\color[HTML]{333333} $72.99_{(39.82)}$} &
  {\color[HTML]{333333} $14.46_{(31.81)}$} &
  {\color[HTML]{333333} $7.86_{(22.51)}$} &
  {\color[HTML]{333333} $77.68_{(37.30)}$} &
  {\color[HTML]{333333} $3.20_{(3.26)}$} &
  {\color[HTML]{333333} $-7.9_{(5.12)}$} &
  {\color[HTML]{333333} $-2.35_{(2.84)}$} \\ \bottomrule
\end{tabular}%
}

\caption{Results of different methods in Mixing game. The first two columns show the average probability of two agents choosing different actions.  The third column displays the individual rewards earned by each agent and the average rewards of the two agents. The values in parentheses correspond to a single standard deviation over trials.}
\label{tab:MixingResults}
\end{table*}

We evaluate the performance of our proposed algorithm, STEP, in three benchmark environments: the Repeated Matrix Game, the Multi-Agent MuJoCo~\cite{ma-mujoco}, and the Highway On-Ramp Merging~\cite{HORM}. 
Appendix contains thorough descriptions of the three environments.
Based on the reward settings (shared or individual rewards) and control types (discrete or continuous action space) in various environments, we compare STEP with various state-of-the-art MARL algorithms.
The main objectives of these experiments are to:
(a) Validate STEP's ability to identify the Stackelberg equilibrium policies;
(b) Evaluate its performance in more challenging cooperative and mixed tasks;
and (c) Assess the effect of the hyperparameter on the its performance.

\subsection{Repeated Matrix Game}

We test STEP using the cooperative matrix game Penalty proposed by Claus and Boutilier~\shortcite{Penalty} and the non-cooperative matrix game Mixing shown in Figure~\ref{fig:matrix}.
Agents receive constant observations at each time step in both games, and the length of repeated matrix games is set to 25.
To cultivate formidable critics, the popular CTDE paradigm utilises combined observation and action to enlarge the information space. However, there is no noticeable performance difference between independent learning (IL) and CTDE for empty state matrix games. 
Therefore, only MAPPO~\cite{MAPPO} and MADDPG~\cite{MADDPG} are considered as baselines for our research.
We present the experimental results over 100 trials of 10000 time steps in Figure~\ref{fig:penalty} and Table~\ref{tab:MixingResults},
which show that STEP outperforms the baselines in both types of matrix games. More detailed results can be found in Appendix.

In Penalty, there are three NE points, and the only SE point corresponds to the optimal NE point.
The challenge lies in the fact that during the exploration phase, 
any deviation from the optimal NE policy by any of the players will result in severe punishment for all players.
The likelihood that agents choose the best NE action decreases as the penalty term increases. 
Consequently,
MAPPO always falls into the trap of penalty and converges to the sub-optimal NE point $(a^1_2,a^2_2)$ at a rate of $100\%$, unless $k=0$.
Similarly, when $k$ declines, MADDPG has a greater probability of converging to the sub-optimal NE point.
Compared to the baselines, SE is the convergence objective for STEP. 
By computing the leader's action, the follower is always able to select the best response action, allowing it to efficiently arrive at the optimal solution regardless of the configuration.
Even under a high penalty of $k=-100$, there is a $93\%$ probability of convergence to the optimal value. 
Due to the use of neural network approximation, modest probability errors are permitted.

\begin{figure}[t]
    \centering
    \includegraphics[width=3.35 in]{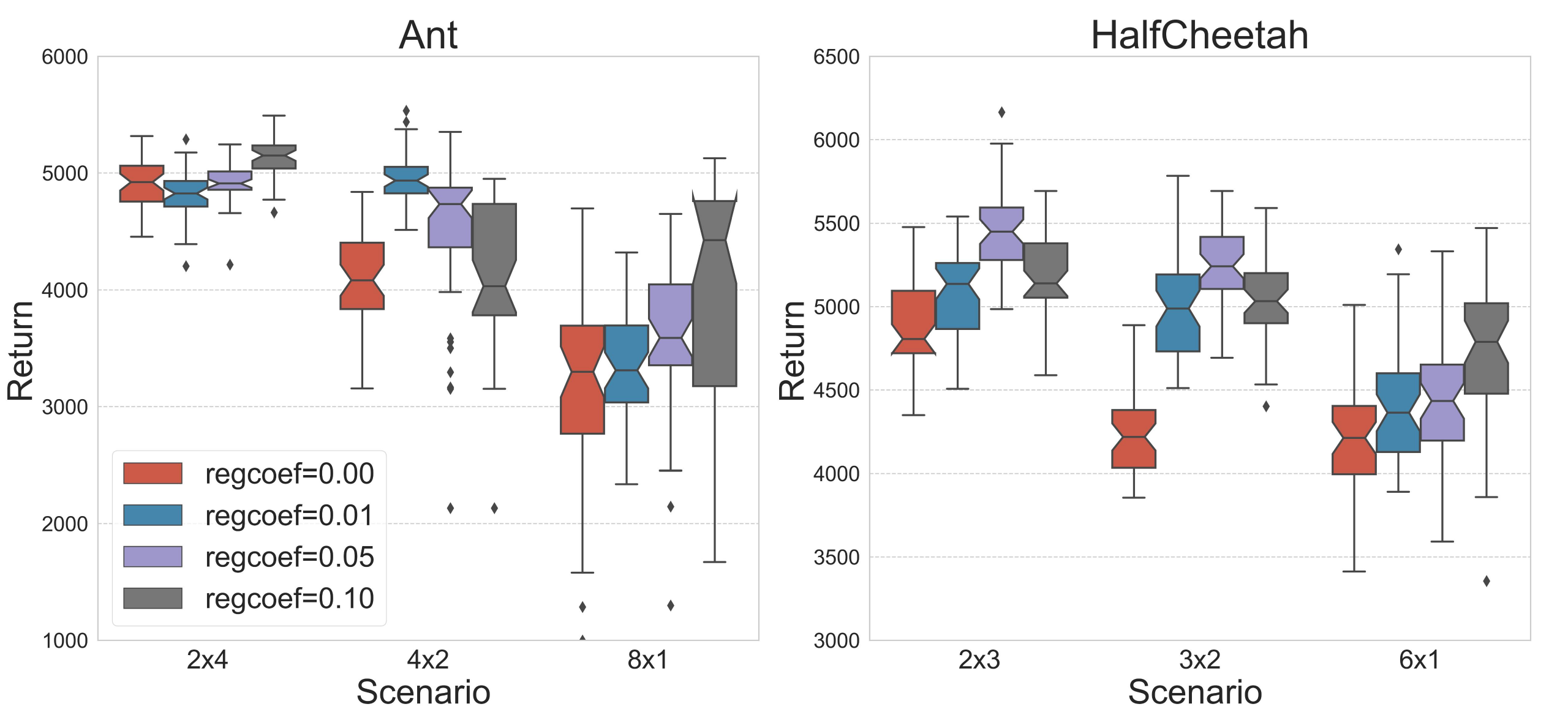}
    \caption{Results of STEP with different regularization coefficients. Note that a larger coefficient produces better results in the case of more agents.}
    \label{fig:regcoef}
\end{figure}

\begin{figure*}[ht]
    \centering
    \includegraphics[width=6.5 in]{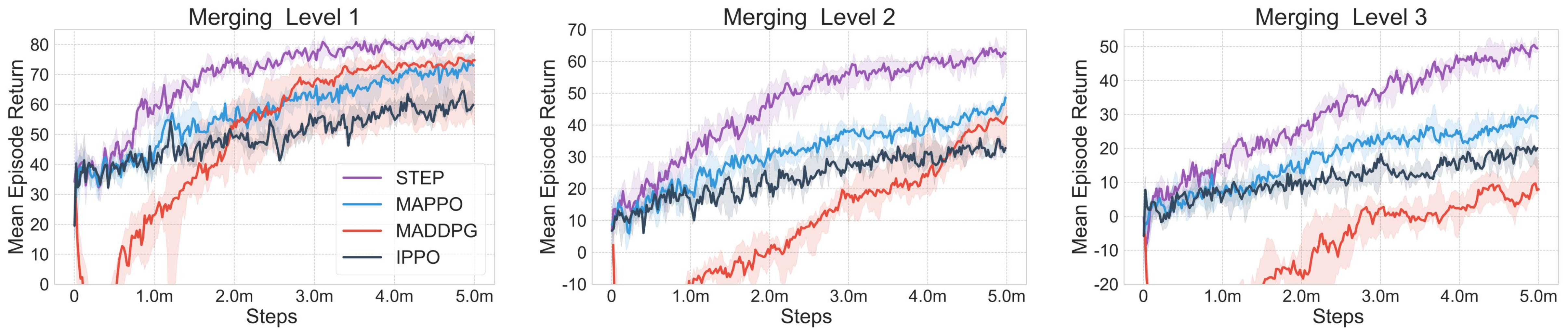}
    \vskip -0.15 in
    \caption{Learning curves for the Highway On-Ramp Merging tasks. Error bars are a 95\% confidence interval across 5 runs.}
    \label{fig:highway_results}
\end{figure*}

Table~\ref{tab:MixingResults} displays the results of different methods in Mixing game. 
We evaluate the agents' coordination based on the average return proposed by Zhang~\shortcite{fully_dencentralized}, as each agent has its own particular reward.
It is clear that there are two optimal solutions in this scenario, with $(a^1_1, a^2_1)$ corresponding to the SE. In addition, the SE is Pareto superior to the only NE.
As expected, STEP is able to rapidly converge to the SE with a probability of nearly $100\%$. In contrast, MAPPO and MADDPG are utterly ineffective and cannot even converge.
It also demonstrates that STEP can effectively solve the problem of selecting multiple optimal solutions to some extent due to the certainty of the SE.

\subsection{Multi-Agent MuJoCo}

\begin{figure}[t]
    \centering
    \includegraphics[width=3.3 in]{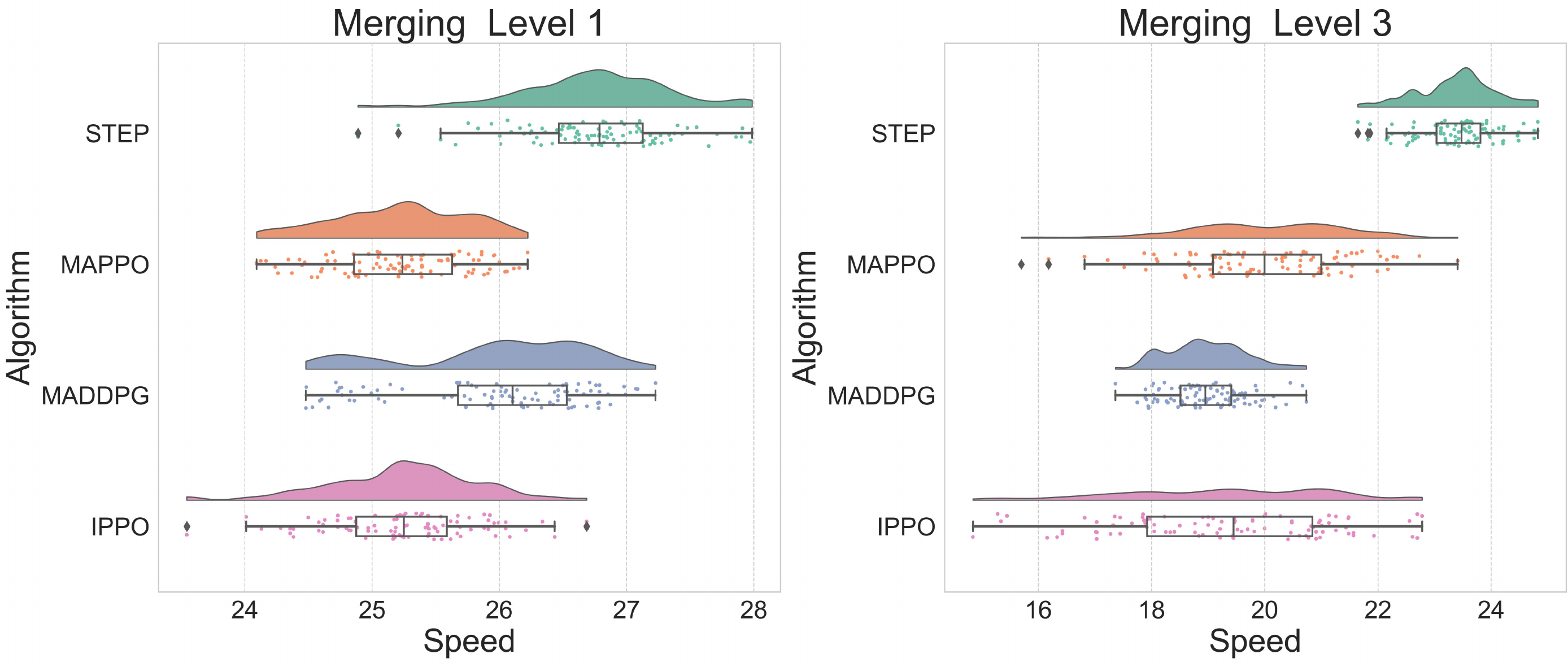}
    \caption{Comparing the testing performance of the average speed among different methods.}
    \label{fig:speed}
\end{figure}
The Multi-Agent MuJoCo environment,
which divides each part of the robot into separate agents that operate each part independently to achieve optimal motion,
is used to assess the performance of the algorithm in a fully cooperative setting,
as well as the impact of the regularization term coefficient on STEP's performance. 
Specifically, 
to measure the collaboration quality of the agents,
we choose representative cooperation scenarios such as Ant and HalfCheetah.
And the results in Many-Agent Ant scenario can be found in Appendix due to the space limit.

Figure~\ref{fig:mujoco_results} demonstrates that STEP outperforms heterogeneous and homogeneous methods in all cases, which is certainly a result of the hierarchical structure among agents.
Notably, the performance gap between STEP and HAPPO narrows as the number of agents increases.
We think it's because STEP achieves the effect of heterogeneous policies through parameter sharing. The number of STEP parameters remains constant regardless of the number of agents, whereas HAPPO with heterogeneous policies builds an individual policy network for each agent.
In this sense, STEP shows a higher capacity to capture and represent diverse and intricate policies.

Figure~\ref{fig:regcoef} depicts the influence of various regularization coefficients on the experimental results.
When the number of agents is small, the performance of various coefficients is similar.
When there are more agents, however, greater coefficients lead to better results. 
The absence of regularization coefficients, on the other hand, has a negative impact on the experimental outcomes, demonstrating the efficiency of the regularization term. This outcome is expected and intuitive.

\subsection{Highway On-Ramp Merging}
Lane merging in high-density, high-speed traffic presents a significant challenge for both autonomous vehicles and human drivers. 
To guarantee safe and efficient merging, vehicles must apply coordination mechanisms that allow them to merge at precise speeds and times. 
In this scenario, each agent has its own goal, and all of them seek to pass through the main road quickly. 
Due to the fact that HAPPO is built for a fully cooperative environment, we only employ MAPPO, IPPO, and MADDPG to compare with STEP in this mixed task scenario. The road conditions vary at different levels.

Results in Figure~\ref{fig:highway_results} show that STEP outperforms the other approaches in the mixed task scenario, with the performance gap between STEP and the other algorithms growing as traffic density and complexity rise. Additionally, the average speed of STEP, as illustrated in Figure~\ref{fig:speed}, is also faster than that of the other algorithms after convergence occurs.

\section{Conclusion}

In this study, we propose a multi-agent reinforcement learning method called STEP, which is inspired by the Stackelberg leadership model, multi-task learning, as well as high-dimensional continuous action decomposition for single agents.
The core insight behind STEP is to make sequential decisions in both temporal and spatial domains that drives MARL to converge to the Stackelberg equilibrium. During the centralized training phase, we construct an efficient interaction mechanism, namely the spatio-temporal sequence Markov game, and utilize a shared N-level policy model to learn policies that aid the agents in achieving efficient coordination during the decentralized execution phase. 
The experimental results demonstrate that our STEP approach achieves state-of-the-art performance.
We believe that asynchronous action coordination and the spatio-temporal sequential decision-making model have further development potential, and how to leverage the advantages of the game structure is a topic worthy of further research.

\bibliographystyle{named}
\bibliography{ijcai23}

\end{document}